\documentclass[10pt, conference]{IEEEtran}

\usepackage[utf8]{inputenc}
\usepackage{amsmath}
\usepackage{amsfonts}

\usepackage{amsthm}
\usepackage{verbatim}
\usepackage{float}

\usepackage{subfigure}
\usepackage{graphicx}
\graphicspath{{Figures/}}
\usepackage{epstopdf}

\usepackage{epsfig}
\usepackage{url}
\usepackage{comment}
\usepackage{cite}
\usepackage{graphics}
\usepackage{color,soul} 
\usepackage{multirow}

\usepackage{todonotes} 

\usepackage{array, blindtext} 
\usepackage{wrapfig}

\usepackage{algorithm}  
\usepackage{algpseudocode}
\usepackage{listings}
\usepackage{psfrag,wrapfig}
\usepackage{paralist} 
\usepackage{subfiles} 

\newtheorem{remark}{Remark}

\newcommand{\eg}{{\it e.g.,}\xspace}
\newcommand{\viz}{{\it viz., }}

\newcommand{\etal}{{\it et~al.}\xspace}
\newcommand{\ie}{{\it i.e.,}\xspace}
\newcommand{\etc}{{\it etc.}}
\newcommand{\ci}{{\it (i) }}
\newcommand{\cii}{{\it (ii) }}

\newcommand{\ca}{{\it (a) }}
\newcommand{\cb}{{\it (b) }}
\newcommand{\cc}{{\it (c) }}
\newcommand{\cd}{{\it (d) }}

\def\QED{\ensuremath{\boldsymbol{\square}}}
\def\markatright#1{\leavevmode\unskip\nobreak\quad\hspace*{\fill}{#1}}
\def\qed{\markatright{\QED}}

\IEEEoverridecommandlockouts
\begin{document}


\title{End-to-End Network Delay Guarantees\\ for Real-Time Systems using SDN}


\author{
\IEEEauthorblockN{Rakesh Kumar\IEEEauthorrefmark{2}, Monowar Hasan\IEEEauthorrefmark{2}, Smruti Padhy\IEEEauthorrefmark{2}\IEEEauthorrefmark{1}\thanks{\IEEEauthorrefmark{1}Smruti Padhy was affiliated with Univesity of Illinois at the time of this work but her affiliation has since changed to Massachussetts Institute of Technology.}, Konstantin Evchenko\IEEEauthorrefmark{2},\\ Lavanya Piramanayagam\IEEEauthorrefmark{6}, Sibin Mohan\IEEEauthorrefmark{2} and Rakesh B. Bobba\IEEEauthorrefmark{4}}
\IEEEauthorblockA{\IEEEauthorrefmark{2}University of Illinois at Urbana-Champaign, USA, 
\IEEEauthorrefmark{6}PES University, India,
\IEEEauthorrefmark{4}Oregon State University, USA}
\IEEEauthorblockA{Email: \IEEEauthorrefmark{2}\{kumar19, mhasan11, evchenk2, sibin\}@illinois.edu, \IEEEauthorrefmark{1}smruti@mit.edu,\\ 
\IEEEauthorrefmark{6}lava281995@gmail.com,
\IEEEauthorrefmark{4}rakesh.bobba@oregonstate.edu} 
}

\maketitle

\thispagestyle{plain}
\pagestyle{plain}

\begin{abstract}

We propose a novel framework that reduces the management and integration
overheads for real-time network flows by leveraging the capabilities
(especially global visibility and management) of software-defined
networking (SDN) architectures. Given the specifications of flows that
must meet hard real-time requirements, our framework synthesizes paths
through the network and associated switch configurations -- to
guarantee that these flows meet their end-to-end timing requirements.
In doing so, our framework makes SDN architectures ``delay-aware'' -- remember
that SDN is otherwise not able to reason about delays. Hence, it
is easier to use such architectures in safety-critical and other latency-sensitive
applications. We demonstrate our principles as well as the feasibility of our approach using
both -- exhaustive simulations as well as experiments using real
hardware switches.


%
\end{abstract}






\section{Introduction}
\label{sec:intro}

Software-defined networking (SDN) \cite{mckeown2008openflow} has become increasingly popular since it
allows for better management of network resources, application of security
policies and testing of new algorithms and mechanisms. It finds use in a wide
variety of domains – from enterprise systems \cite{levin2013incremental} to cloud computing services \cite{jain2013network}, from
military networks \cite{spencer2016towards} to power systems \cite{pfeiffenberger2014evaluation} \cite{aydeger2016software}. The global view of the
network obtained by the use of SDN architectures provides significant advantages when compared
to traditional networks. It allows designers to push down rules to the various
nodes in the network that can, to a fine level of precision, manage the
bandwidth and resource allocation for flows through the entire network. However, current SDN architectures do not reason about delays. On the other hand, real-time systems (RTS), especially those with stringent timing constraints,
need to reason about delays. {\em Packets must be delivered between hosts with
guaranteed upper bounds on end-to-end delays}. Examples of such systems include
avionics, automobiles, industrial control systems, power substations,
manufacturing plants, \etc~ 

While RTS can include different types of traffic\footnote{For instance, 
\ca high priority/criticality traffic that is essential for the
correct and safe operation of the system; 
\cb medium criticality traffic that is critical to the correct operation
of the system, but with some tolerances in delays, packet drops, \etc; and 
\cc low priority traffic -- essentially all other traffic in the system 
that does not really need guarantees on delays or bandwidth such as engineering
traffic in power substations, multimedia flows in aircraft, \etc}, in this
paper we focus on the high priority flows that have stringent timing
requirements, predefined priority levels and can tolerate little to no loss of
packets.  We refer to such traffic as ``Class I'' traffic.  Typically, in many
safety-critical RTS, the properties of all Class I flows are well known, \ie
designers will make these available ahead of time. Any changes
(addition/removal of flows or modifications to the timing or bandwidth
requirements) will often require a serious system redesign. The number (and
properties) of other flows could be more dynamic -- consider the on-demand
video situation in an airplane where new flows could arise and old ones stop based on the
viewing patterns of passengers. 

Current safety-critical systems often have separate networks (hardware and
software) for each of the aforementioned types of flows (for safety and
sometimes security reasons).  This leads to significant overheads (equipment,
management, weight, \etc) and also potential for errors/faults and even
increased attack surface and vectors.  Existing systems, \eg avionics
full-duplex switched Ethernet (AFDX) \cite{ARINC2009,land2009,Charara2006},
controller area network (CAN) \cite{can2014}, \etc~that are in use in many of
these domains are either proprietary, complex, expensive and might even require
custom hardware. Despite the fact that AFDX switches ensure timing determinism,
packets transmitted on such switches may be changed frequently at run-time when
sharing resources (\eg bandwidth) among different networks
\cite{afdx_openflow}. In such situations, a dynamic configuration is required
to route packets based on switch workloads and flow delays to meet all the high
priority Quality of Service (QoS) requirements (\eg end-to-end delay). In
addition AFDX protocols require custom hardware \cite{afdx_evolution}.

In this paper we present mechanisms to {\em guarantee end-to-end delays for
high-criticality flows (Class I) on networks constructed using SDN switches.}
The advantage of using SDN is that it provides a centralized mechanism for
developing and managing the system. The global view is useful in providing the
end-to-end guarantees that are required. Another advantage is that the
hardware/software resources needed to implement all of the above types of
traffic can be reduced since we can use the same network infrastructure
(instead of separate ones as is the case currently).  On the other hand, the
current standards used in traditional SDN (OpenFlow \cite{mckeown2008openflow,
specification20131}) generally do not support end-to-end delay guarantees or
even existing real-time networking protocols such as AFDX.  Retrofitting
OpenFlow into AFDX is not straightforward and is generally less effective
\cite{deterministic_openflow}.

A number of issues arise while developing a software-defined networking
infrastructure for use in real-time systems. For instance, Class I flows 
need to meet their \textit{timing} (\eg end-to-end delay) requirements
for the real-time system to function correctly. Hence, we
need to \textit{find a path} through the network, along with necessary resources, that will meet these
guarantees. However, current SDN implementations reason about resources like bandwidth instead of delays. Hence, we must find a way to extend the SDN infrastructure to reason about delays for use in RTS. Further, in contrast to traditional SDNs, it is not necessary to find the
\textit{shortest} path through the network. Oftentimes, Class I flows can
arrive \textit{just in time} \cite{Qian2015, Oh2015}, \ie just before their
deadline -- there is no real advantage in getting them to their destinations
well ahead of time. Thus, path layout for real-time SDN is a \textit{non-trivial}
problem since, \ci we need to understand the delay(s) caused by individual
nodes (\eg switches) on a Class I flow and \cii compose them along 
the delays/problems caused by the presence of other flows in
that node as well as the network in general.

In this work\footnote{A preliminarily version of the work is under submission to the 2017 RTN workshop that does not have published proceedings. In this paper we extend the workshop version with more comprehensive experiments (Section \ref{sec:evaluation}) and evaluation on actual hardware switches (Section \ref{sec:motivating_example}).} we consider Class I (\ie high-criticality) flows and develop a
scheme to meet their timing constraints\footnote{We will work on integrating
other types of traffic in future work.}. 
We evaluate the effectiveness of the proposed approach with various custom
topology and UDP traffic (Section \ref{sec:evaluation}). The main contributions of this work are summarized as follows:

\begin{enumerate}
\item We developed mechanisms to guarantee timing constraints for traffic in hard real-time
	systems based on COTS SDN hardware (Sections
	\ref{sec:system_model}, \ref{sec:approach} and \ref{sec:algo}).

\item We illustrate the requirements for isolating flows into different queues
	to provide stable quality of experience in terms of end-to-end delays
	(Section \ref{sec:motivating_example}) even in the presence of other
	types of traffic in the system.
\end{enumerate}

\section{Background}
\label{sec:background}

\subsubsection{The Software Defined Networking Model}

\begin{figure}[ht!]
\centering
\vspace*{-1em}
\includegraphics[width=2.2in]{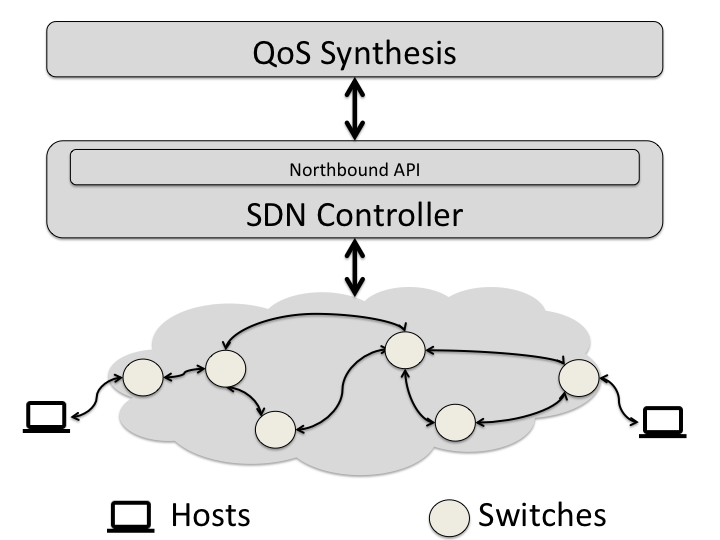}
\caption{An SDN with a six switch topology. Each switch also connects to the controller via a management port (not shown). The QoS Synthesis module (Section \ref{sec:implementation}) synthesizes flow rules by using the northbound API.}
\label{fig:sdn_model}
\vspace{-0.2in}
\end{figure}

In traditional networking architectures, control and data planes coexist on
network devices. SDN architectures simplify access the system by logically centralizing the control-plane state in a
\textit{controller} (see Figure~\ref{fig:sdn_model}). This programmable and
centralized state then drives the network devices that perform homogeneous
forwarding plane functions \cite{casado2009rethinking} and can be modified to
control the behavior of the SDN in a flexible manner. 

In order to construct a logically centralized \textit{state} of the SDN system, the
controller uses management ports to gauge the current topology and
gather data-plane state from each switch. This state is then made available
through a \textit{northbound} API to be used by the applications. An
application (\eg our prototype proposed in this paper) uses this API to
obtain a snapshot of the SDN state. 
This state also includes the network topology.

\subsubsection{The Switch}

An SDN switch consists of table processing pipeline and a collection of
physical \textit{ports}. Packets arrive at one of the ports and they are
processed by the pipeline made up of one or more flow tables. Each flow table
contains \textit{flow rules} ordered by their priority. Each flow rule
represents an atomic unit of decision-making in the control-plane. During the
processing of a single packet, \textit{actions} (\eg decision-making entities)
can modify the packet, forward it out of the switch or drop it. 

When a packet arrives at a switch, it is compared with flow rules in one or
more flow table pipelines. In a given table, the contents of the packet header
are compared with the flow rules in decreasing order of rule priority. When
a matching flow rule is found, the packet is assigned a set of
actions specified by the flow rule to be applied at the end of table processing
pipeline. Each flow rule comprises of two parts:

\begin{itemize}
\item \textbf{Match}: is set of packet header field values that a given flow rule applies to. Some are characterized by single values (\eg VLAN ID: 1, or TCP Destination Port: 80), others by a range (\eg Destination IP Addresses: 10.0.0.0/8). If a packet header field is not specified then it is considered to be a wild card.

\item \textbf{Instructions Set}: is a set of actions applied by the flow rule to a matching packet. The actions can specify the egress port ($\mathsf{Output Port}$) for packets matching the rule. Furthermore, in order to make the appropriate allocation of bandwidth for the matching packets, the OpenFlow \cite{specification20131} specification  provides two mechanisms:

\begin{itemize}

\item \textbf{Queue References}: Every OpenFlow switch is capable of providing isolation to traffic from other flows by enqueuing them on separate queues on the egress port. Each queue has an associated QoS configuration that includes, most importantly, the service rate for traffic that is enqueued in it. The OpenFlow standard itself does not provide mechanisms to configure queues, however, each flow rule can refer to a specific queue number for a port, besides the $\mathsf{Output Port}$.

\item \textbf{Meters}: Beyond the isolation provided by using queues, OpenFlow switches are also capable of limiting the rate of traffic in a given network flow by using objects called meters. The meters on a switch are stored in a meter table and can be added/deleted by using messages specified in OpenFlow specification. Each meter has an associated metering rate. Each flow rule can refer to only a single meter. 

\end{itemize}

\end{itemize}

\section{System Model}
\label{sec:system_model}

Consider an SDN topology ($N$) with open flow switches and controller and a set
hrof real-time flows ($F$) with specified delay and bandwidth guarantee
requirements. The {\em problem is to find paths for the flows (through the 
topology) such that the flow requirements (\ie end-to-end delays) 
can be guaranteed for the maximum number of critical flows}.
We model the network as an undirected graph $N(V,E)$ where $V$ is
the set of nodes, each representing a switch port in a given network and $E$ is
set of the edges\footnote{We use the terms \textit{edge} and \textit{link}
interchangeably throughput the paper.}, each representing a possible path for
packets to go from one switch port to another.  Each port $v \in V$ has a set
of queues $v_q$ associated with it, where each queue is assigned a fraction of
bandwidth on the edge connected to that port.  

Consider a set $F$ of unidirectional, real-time flows that require delay and bandwidth
guarantees. The flow $f_k \in F$ is given by a four-tuple $(s_k, t_k,
D_k, B_k)$, where $s_k \in V$ and $t_k \in V$ are ports (the source 
and destination respectively) in the graph, $D_k$ is
the maximum delay that the flow can tolerate and $B_k$ is the maximum required
bandwidth by the flow. We assume that flow priorities are distinct and the
flows are prioritized based on a \textit{``delay-monotonic''} scheme \viz
the end-to-end delay budget represents higher priority (\ie $pri(f_i) > pri(f_j)$ if
$D_i < D_j,~ \forall f_i, f_j \in F$ where $pri(f_k)$ represents priority of
$f_k$). 

For a flow to go from the source port $s_k$ to a destination port $t_k$, it
needs to traverse a sequence of edges, \ie a flow path $\mathcal{P}_k$. The
problem then, is to synthesize flow rules that use queues at each edge $(u,v) \in
\mathcal{P}_k$ that can handle \textit{all} flows $F$ in the given system
while still meeting each flow's requirement. If $d_{f_k}(u,v)$ and $b_{f_k}(u,v)$ is
the delay faced by the flow and bandwidth assigned to the flow at each edge
$(u,v) \in E$ respectively, then $\forall f_k \in F$ and $\forall (u,v) \in
\mathcal{P}_k$ the following constraints need to be satisfied:

\begin{align}
\sum_{(u,v)\in \mathcal{P}_k} d_{f_k}(u,v) &\leq D_k, \quad \forall f_k \in F  \label{eq:con_delay}\\
b_{f_k}(u,v) &\geq B_k, \quad \forall (u,v) \in \mathcal{P}_k, \forall f_k \in F. \label{eq:con_bw}
\end{align}

This problem needs to be solved at two levels:
\begin{itemize}
\item \textit{Level 1}: Finding the path layout for each flow such that it
	satisfies the flows' delay and bandwidth constraints. We formulate this
	problem as a multi-constrained path (MCP) problem and describe the
	solution in Sections \ref{sec:approach} and \ref{sec:algo}.
\item \textit{Level 2}: Mapping the path layouts from Level 1 on to the
	network topology by using the mechanisms available in OpenFlow. We
	describe details of our approach in Section~\ref{sec:implementation}.
\end{itemize}


In addition to the aforementioned delay and bandwidth constraints (see Eqs.~(\ref{eq:con_delay}) and (\ref{eq:con_bw})), we need to map flows assigned to a port to the queues at the port. Two possible
approaches are: 
\ca {\em allocate each flow to an individual queue} or 
\cb {\em multiplex flows onto a smaller set of queues} and dispatch the packets based on priority. In fact, as we illustrate in the following section, the queuing approach used will impact the delays faced by the flows at each link. Our intuition is that
the {\em end-to-end delays are lower and more stable} when {\em separate queues}
are provided to each critical flow -- especially as the rates for the flows
get closer to their maximum assigned rates. Given the deterministic nature of many RTS, the number of critical flows are often limited and well defined (\eg known at design time). Hence, such over-provisioning is an acceptable design choice -- from computing power to network resources (for instance one queue per critical real time flow). We carried out some experiments 
to demonstrate this (and to highlight the differences between these two strategies) 
-- this is outlined in the following section. 







\subsection{Queue Assignment Strategies}
\label{sec:motivating_example}

\begin{figure*}[htb]
\vspace{-1em}
\addtolength{\subfigcapskip}{-0.2in}
\centerline{\subfigure[]{\includegraphics[width=3in]{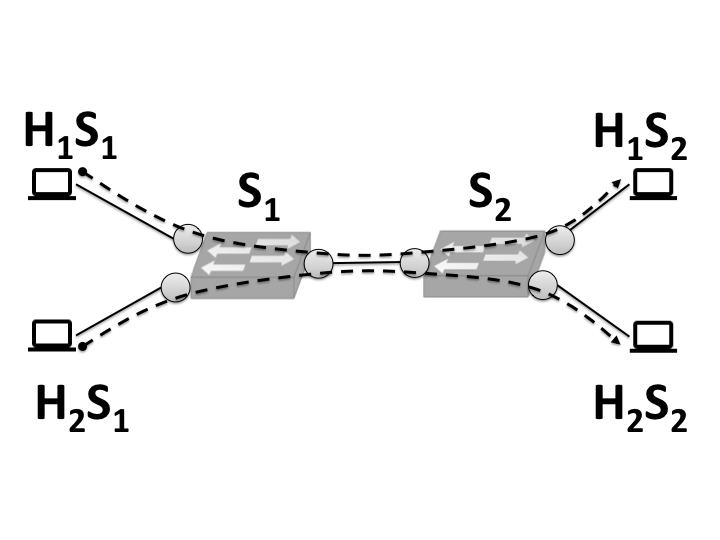}
\label{fig:two_switch_topology_with_flows}}
\hfil
\hfill
\vspace*{-1em}
\subfigure[]{\includegraphics[width=3.5in]{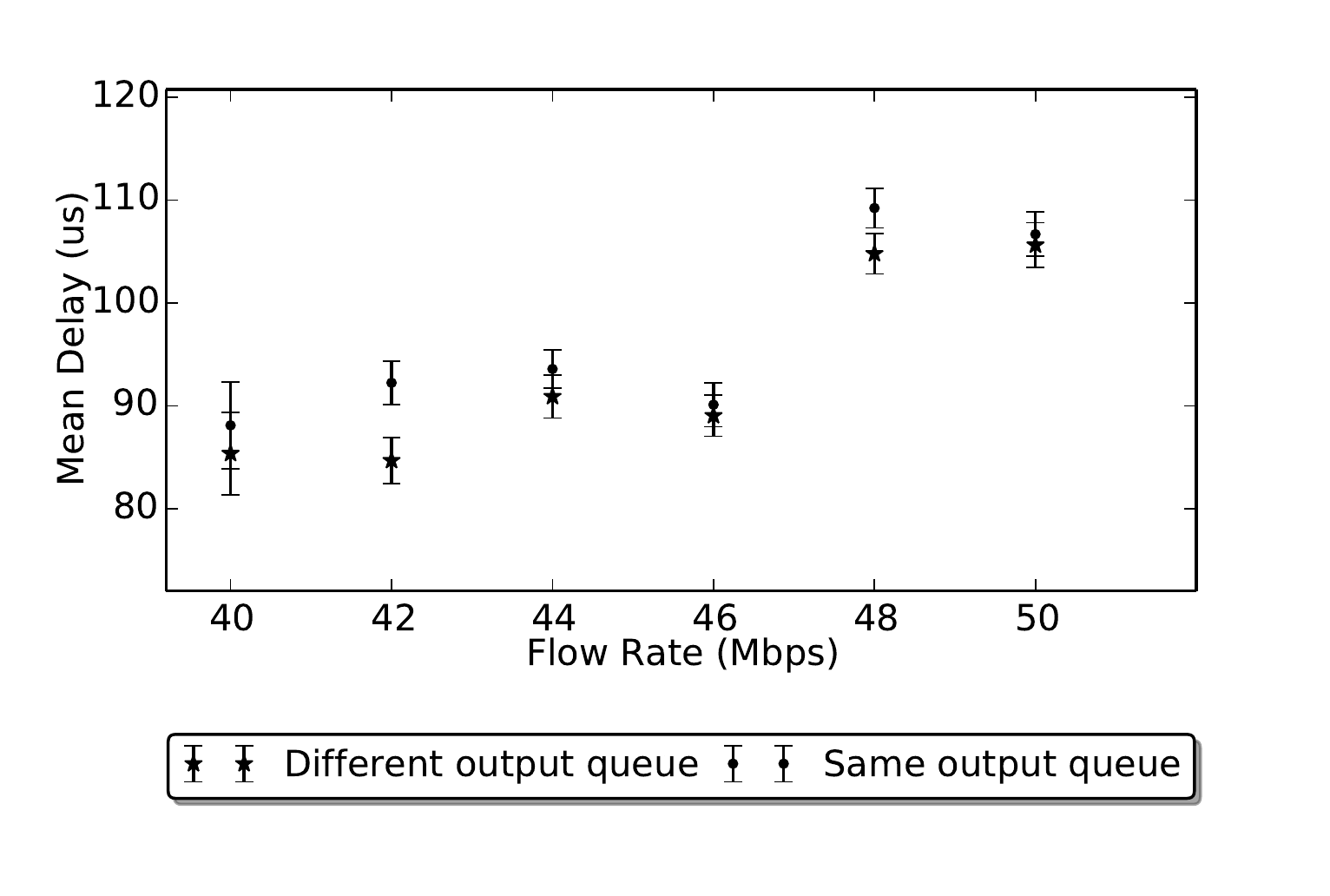}
\label{fig:delay_measurements}}\hfil}
\caption{Delay measurement experiments: (a) The two-switch, four host topology used in the experiments with the active flows. (b) The measured mean and $99^{th}$ percentile per-packet delay for the packets in the active flows in 30 iterations.}
\label{fig:motivating_example_figure}
\vspace{-0.2in}
\end{figure*}

We intend to synthesize configurations for Class I traffic such that it ensures
{\em complete isolation of packets for each designated class I flow}. 

In order to test how using output queues can provide isolation to flows
in a network so that each can meet its delay and bandwidth requirements
simultaneously, we performed experiments using OpenFlow enabled hardware switches\footnote{We also conduct similar experiments with software simulations (\eg by using Mininet \cite{lantz2010network} topology) and observe similar trends (see Appendix \ref{appsec:mininet_simple_topo}).}. The experiments use a simple topology that contains two white box Pica8 P-3297 \cite{pica8} switches (\texttt{s1},
\texttt{s2}) connected via a single link as shown in Figure
\ref{fig:motivating_example_figure}(a). Each switch has two hosts connected to
it. Each host is a Raspberry Pi 3 Model B \cite{raspberry} running Raspbian Linux. 

We configured flow rules and queues in the switches to enable connectivity
among hosts at one switch with the hosts at other switch. We experimented with
two ways to queue the packets as they cross the switch-to-switch link:
\ci in one case, we queue packets belonging to the two flows {\em separately} in two queues
(\ie each flow gets its own queue), each configured at a maximum rate of 50 Mbps
\cii in the second case, we queue packets from both flows in the {\em same queue}
configured at a maximum rate of 100 Mbps. 

After configuring the flow rules and queues, we used 
\texttt{netperf} \cite{netperf} to generate following packet flows: the first starting at the host \texttt{h1s1} destined to
host \texttt{h1s2} and the second starting at host \texttt{h2s1} with a destination 
host \texttt{h2s2}. Both flows are identical and are triggered simultaneously to last for 
15 seconds. We changed the rate at which the traffic is sent across both flows to measure the average per-packet delay. Figure \ref{fig:motivating_example_figure}(b) plots the average value
and standard error over 30 iterations. The x-axis indicates the rate at which
the traffic is sent via \texttt{netperf}, while the y-axis shows the average per-packet delay. The following key observations stand out:
\begin{enumerate}
	\item  The per-packet average delay increases in both cases
	as traffic send rate approaches the configured rate of 50 Mbps. This is
	an expected queue-theoretic outcome and motivates the need for slack
	allocations for all applications in general. For example, if an
	application requires a bandwidth guarantee of 1 Mbps, it should be
	allocated 1.1 Mbps for minimizing jitter.
\item The case with separate queues experiences lower average per-packet delay when flow rates approach the
	maximum rates. This indicates that when more than one flow
	uses the same queue, there is interference caused by both flows to
	each other. This becomes a source of unpredictability and eventually may
	cause the end-to-end delay guarantees for the flows to be not met or perturbed
	significantly.
\end{enumerate}

Thus, {\em isolating flows using separate queues results in lower and more stable
delays} especially when traffic rate in the flow approaches the configured maximum rates.
The maximum delay along a single link can be measured. Such measurements can
then be used as input to a path allocation algorithm that we describe in the following section.

\section{Path Layout: Overview and Solution}
\label{sec:approach}



We now present a more detailed version of the problem (composing
paths that meet end-to-end delay constraints for critical real-time flows) and
also an overview of our solution.

\subsubsection*{Problem Overview}

Let $\mathcal{P}_k$ be the path from $s_k$ to $t_k$ for flow $f_k$ that needs
to be determined. Let $\mathfrak{D}(u, v)$ be the delay incurred on the edge
$(u,v) \in E$. 
The total delay for $f_k$ over the path $\mathcal{P}_k$ is  given by
\begin{equation}
\mathfrak{D}_k(\mathcal{P}_k) = \sum_{(u,v) \in \mathcal{P}_k} \mathfrak{D}(u, v).
\end{equation}
Therefore we define the following constraint on end-to-end delay for the flow $f_k$ as
\begin{equation} \label{eq:delay_con}
\mathfrak{D}_k(\mathcal{P}_k) \leq D_k.
\end{equation}


Note that the end-to-end delay for a flow over a path has following delay components: 
\ca processing time of a packet at a switch,
\cb propagation on the physical link,
\cc transmission of packet over a physical link, and
\cd queuing at the ingress/egress port of a switch.
As discussed in the Section \ref{sec:system_model}, we use separate queues for
each flow with assigned required rates. We also overprovision the bandwidth
for such flows so that critical real-time flows do not experience queueing delays.
Hence, we consider queuing delays to be negligible. We discuss how to obtain 
the values of other components of delay in Appendix \ref{subsec:delay_cal}.

The second constraint that we consider in this work is {\em bandwidth utilization}, that
for an edge $(u,v)$ for a flow $f_k$, can be defined as:  
\begin{equation}
\mathfrak{B}_{k}(u,v) = \frac{B_k}{B_e(u,v)}
\end{equation}
where $B_k$ is the bandwidth requirement of $f_k$ and $B_e(u,v)$ is total bandwidth of an edge $(u,v) \in E$.
Therefore, bandwidth utilization over a path ($\mathcal{P}_k$), for a flow $f_k$ is defined as:  
\begin{equation}
	\mathfrak{B}_k(\mathcal{P}_k) = \sum_{(u,v)\in \mathcal{P}_k} \mathfrak{B}_{k}(u,v).
\end{equation}
Note that the bandwidth utilization over a path $\mathcal{P}_k$ for flow  $f_k$ is bounded by
\begin{equation}
  \mathfrak{B}_k(\mathcal{P}_k) \leq \max_{(u,v) \in E} \mathfrak{B}_{k}(u,v) |V|.
\end{equation}
where $|V|$ is the cardinality of a set of nodes (ports) in the topology $N$.
Therefore in order to ensure that the bandwidth requirement $B_k$ of the flow $f_k$ is guaranteed, it suffices to consider the following constraint on bandwidth utilization
\begin{equation} \label{eq:bw_con}
\mathfrak{B}_k(\mathcal{P}_k) \leq \widehat{B}_k
\end{equation}
where $\widehat{B}_k = \max\limits_{(u,v) \in E} \mathfrak{B}_{k}(u,v) |V|$
\begin{remark}
The selection of an optimal path for each flow $f_k \in F$ subject to delay and
bandwidth constraints in Eq. (\ref{eq:delay_con}) and (\ref{eq:bw_con}),
respectively can be formalized as a multi-constrained path (MCP) problem that
is known to NP-complete \cite{mcp_np_hard1}. 
\end{remark}

Therefore we extend a polynomial-time heuristic similar to that presented in
literature\cite{mcp_klara}. The key idea is to \textit{relax}
one constraint (\eg delay or bandwidth) at a time and try to obtain a solution.
If the original MCP problem has a solution, one of the relaxed versions of the
problem will also have a solution \cite{mcp_klara}. In what follows, we briefly
describe the polynomial-time solution for the path layout problem.

\subsubsection*{Polynomial-time Solution to the Path Layout Problem}

Let us represent the delay and bandwidth constraint as follows
\begin{align}
\widetilde{\mathfrak{D}}_k(u,v) &= \left\lceil \frac{X_k \cdot \mathfrak{D}(u,v)}{D_k} \right\rceil \label{eq:delay_relax} \\
\widetilde{\mathfrak{B}}_k(u,v) &= \left\lceil \frac{X_k \cdot \mathfrak{B}_k(u,v)}{\widehat{B}_k} \right\rceil \label{eq:bw_relax}
\end{align}
where $X_k$ is a given positive integer. For instance, if we relax the bandwidth constraint (\eg represent $\mathfrak{B}_k(\mathcal{P}_k)$ in terms of $\widetilde{\mathfrak{B}}_k(\mathcal{P}_k) = \sum_{(u,v)\in \mathcal{P}_k} \widetilde{\mathfrak{B}}_{k}(u,v)$), Eq. (\ref{eq:bw_con}) can be rewritten as
\begin{equation}
\widetilde{\mathfrak{B}}_k(\mathcal{P}_k) \leq X_k.
\end{equation}
Besides, the solution to this relaxed problem will also be a solution to the
original MCP \cite{mcp_klara}. Likewise, if we relax the delay constraint, Eq.
(\ref{eq:delay_con}) can be rewritten as 
\begin{equation}
\widetilde{\mathfrak{D}}_k(\mathcal{P}_k) = \sum_{(u,v)\in \mathcal{P}_k} \widetilde{\mathfrak{D}}_{k}(u,v) \leq X_k.
\end{equation}

Let the variable $d_k[v, i]$ preserve an \textit{estimate} of the path from
$s_k$ to $t_k$ for $\forall v \in V$, $i \in \mathbb{Z}^+$ (refer to
Algorithm \ref{alg:mcp_org}). There exists a solution (\eg a path
$\mathcal{P}_k$ from $s_k$ to $t_k$) if \textit{any} of the two conditions is
satisfied when the \textit{original MCP problem is solved by the heuristic}.
\begin{itemize}
\item \textit{When the bandwidth constraint is relaxed:} The delay and (relaxed) bandwidth constraints, \eg $\mathfrak{D}_k(\mathcal{P}_k) \leq D_k$ and $\widetilde{\mathfrak{B}}_k(\mathcal{P}_k) \leq X_k$ are satisfied if and only if $$d_k[t, i] \leq D_k, ~~ \exists i \in [0, X_k] \wedge i \in \mathbb{Z}.$$

\item \textit{When the delay constraint is relaxed:} The (relaxed) delay and bandwidth constraints, \eg $\widetilde{\mathfrak{D}}_k(\mathcal{P}_k) = \sum_{(u,v)\in \mathcal{P}_k} \widetilde{\mathfrak{D}}_{k}(u,v) \leq X_k$ and $\mathfrak{B}_k(\mathcal{P}_k) \leq \widehat{B}_k$ are satisfied if and only if $$d_k[t, i] \leq X_k, ~~ \exists i \in [0, \widehat{B}_k] \wedge i \in \mathbb{Z}.$$
\end{itemize}



\section{Algorithm Development} 
\label{sec:algo}

\subsection{Path Layout}

Our proposed approach is based on a polynomial-time solution to the MCP problem presented in literature \cite{mcp_klara}. Let us consider $\mathtt{MCP\_HEURISTIC}$($N, s, t, W_1, W_2, C_1, C_2$), an instance of polynomial-time heuristic solution to the MCP problem that finds a path $\mathcal{P}$ from $s$ to $t$ in any network $N$, satisfying constraints $W_1(\mathcal{P}) \leq C_1$ and $W_2(\mathcal{P}) \leq C_2$. 


The heuristic solution of MCP problem, as summarized in Algorithm \ref{alg:mcp_org} works as follows. Let
\begin{equation}
\Delta(v,i) = \min\limits_{\mathcal{P} \in P(v,i)} W_1(\mathcal{P})
\end{equation}
where $P(v,i) = \lbrace \mathcal{P} ~|~ W_2(\mathcal{P}) = i, \mathcal{P} \text{ is any path from } s \text{ to } t \rbrace$ is the smallest $W_1(\mathcal{P})$ of those paths from $s$ to $v$ for which $W_2(\mathcal{P}) = C_2$. For each node $v \in V$ and each integer $i \in [0, \cdots, C_2]$ we maintain a variable $d[v, i]$ that keeps an estimation of the smallest $W_1(\mathcal{P})$. The variable initialized to $+ \infty$ (Line 3), which is always greater than or equal to $\delta(v,i)$. As the algorithm executes, it makes better estimation and eventually reaches $\Delta(v,i)$ (Line 8-15). 
Line 3-17 in Algorithm \ref{alg:mcp_org} is similar to the single-cost path selection approach presented in earlier work \cite[Sec. 2.2]{mcp_klara} and for the purposes of this work, 
we have extended the previous approach for our formulation.

We store the path in the variable $\pi[v,i], \forall v \in V, \forall i \in [0, \cdots, C_2]$. When the algorithm finishes the search for path (Line 17), there will  be a solution if and only if the following condition is satisfied \cite{mcp_klara} 
\begin{equation} \label{eq:path_cond}
\exists i \in [0, \cdots, C_2],~~ d[t,i] \leq C_1.
\end{equation}
If it is not possible to find any path (\eg the condition in Eq. (\ref{eq:path_cond}) is not satisfied), the algorithm returns False (Line 41). If there exists a solution (Line 19), we extract the path by backtracking (Line 21-29).
Notice that the variable $\pi[v, i]$ keeps the immediate preceding node of $v$ on the path (Line 13). Therefore, the path can be recovered by tracking $\pi$ starting from destination $t$ through all immediate nodes until reaching the source $s$. 
Based on this MCP abstraction, we developed a path selection scheme  considering delay and bandwidth constraints (Algorithm \ref{alg:mcp_path_delay_bw}) that works as follows.

\begin{algorithm}[t]

\newcommand{\algorithmicbreak}{\textbf{break}}
\newcommand{\Break}{\State \algorithmicbreak}
\renewcommand\algorithmiccomment[1]{%
 {\it /* {#1} */} %
}
\renewcommand{\algorithmicrequire}{\textbf{Input:}}
    \renewcommand{\algorithmicensure}{\textbf{Output:}}
    
\renewcommand{\algorithmicforall}{\textbf{for each}}

\begin{algorithmic}[1]
\begin{footnotesize}
\Require The network $N (V,E)$, source $s$, destination $t$, constraints on links $W_1 = [w_1(u,v)]_{\forall(u,v) \in E}$ and $W_2 = [w_2(u,v)]_{\forall(u,v) \in E}$, and the bounds on the constraints $C_1  \in \mathbb{R}^+$ and $C_2 \in \mathbb{R}^+$ for the path from $s$ to $t$.
   \Ensure The path $\mathcal{P}^*$ if there exists a solution (\eg $W_1(\mathcal{P^*}) \leq C_1$ and $W_2(\mathcal{P^*}) \leq C_2$), or $\mathsf{False}$ otherwise.

\vspace{0.8em}

\Function{$\mathtt{MCP\_HEURISTIC}$}{$N, s, t, W_1, W_2, C_1, C_2$}
    \State \Comment{Initialize local variables}

    \State $d[v, i] := \infty, \pi[v, i] := NULL$, ~ $\forall v \in V$,~ $\forall i \in [0, C_2] \wedge i \in \mathbb{Z}$
    \State $d[s, i] := 0$~ $\forall i \in [0, C_2] \wedge i \in \mathbb{Z}$
    \State \Comment{Estimate path}
      \For{$i \in |V|-1$}
    	\ForAll{$j \in [0,C_2] \wedge j \in \mathbb{Z}$}
    		\ForAll{edge $(u,v) \in E$}
    			\State $j' := j + w_2(u,v)$
                \If {$j' \leq C_2$ \textbf{ and } $d[v, j'] > d[u, j] + w_1(u,v)$}
	               \State \Comment{Update estimation}
                	\State $d[v, j'] := d[u, j] + w_1(u,v)$  
                    \State $\pi[v, j'] := u$ ~\Comment{Store the possible path}
                \EndIf
	    	\EndFor
	    \EndFor
    \EndFor
    \State \Comment{Check for solution}
    \If{$d[t, i] \leq C_1 \textrm{ for } \exists i \in [0, C_2] \wedge i \in \mathbb{Z}$ }
    \State \Comment{Solution found, obtain the path by backtracking}
    
    \State $\mathcal{P} := $ \O, $done := \mathsf{False}$, $currentNode := t$ 
    \State \Comment{Find the path from $t$ to $s$}
    
    \While{\textbf{not} $done$} 
    	\ForAll{$j \in [0,C_2] \wedge j \in \mathbb{Z}$}
        	\If {$\pi[currentNode, j]$ \textbf{not}  $NULL$}                    			\State  
            add $currentNode$ to $\mathcal{P}$
            \If {$currentNode = s$} 
            \State $done := \mathsf{True}$ \Comment{Backtracking complete}
            \Break
            \EndIf
            \State \Comment{Search for preceding hop}
			\State $currentNode := \pi[currentNode, j]$ 
            \Break
            \EndIf

        \EndFor
    \EndWhile
    \State \Comment{Reverse the list to obtain a path from $s$ to $t$}
    \State $\mathcal{P}^*$ := \textrm{reverse}($\mathcal{P}$) 
    \State \Return $\mathcal{P}^*$
    \Else
    \State \Return $\mathsf{False}$ \Comment{No Path found that satisfies $C_1$ and $C_2$}
    \EndIf

\EndFunction

\end{footnotesize}

\end{algorithmic}
\caption{Multi-constraint Path Selection}
\label{alg:mcp_org}
\end{algorithm}

\begin{algorithm}[ht]

\newcommand{\algorithmicbreak}{\textbf{break}}
\newcommand{\Break}{\State \algorithmicbreak}
\renewcommand\algorithmiccomment[1]{%
 {\it /* {#1} */} %
}
\renewcommand{\algorithmicrequire}{\textbf{Input:}}
    \renewcommand{\algorithmicensure}{\textbf{Output:}}
    
\renewcommand{\algorithmicforall}{\textbf{for each}}

\begin{algorithmic}[1]
\begin{footnotesize}
\Require The network $N (V,E)$, set of flows $F$, delay and bandwidth utilization constraints on links  $\mathfrak{D}_k = [\mathfrak{D}_k(u,v)]_{\forall(u,v) \in E},~ \widetilde{\mathfrak{D}}_k = [\widetilde{\mathfrak{D}}_k(u,v)]_{\forall(u,v) \in E}$ and $\mathfrak{B}_k = [\mathfrak{B}_k(u,v)]_{\forall(u,v) \in E},~ \widetilde{\mathfrak{B}}_k = [\widetilde{\mathfrak{B}}_k(u,v)]_{\forall(u,v) \in E}$, for each flow $f_k \in F$, respectively, and the delay and bandwidth bounds $D_k  \in \mathbb{R}^+$ and $\widehat{B}_k \in \mathbb{R}^+$, respectively, and positive constant $X_k \in \mathbb{Z}$, $\forall f_k \in F$.
   \Ensure The path vector $\boldsymbol{\mathcal{P}} = [\mathcal{P}_k]_{\forall f_k \in F}$  where $\mathcal{P}_k$ is the path if the delay and bandwidth constraints (\eg $\mathfrak{D}_k(\mathcal{P}_k) \leq D_k$ and $\mathfrak{B}_k(\mathcal{P}_k) \leq \widehat{B}_k$) are satisfied for $f_k$, or $\mathsf{False}$ otherwise. 
\vspace{0.2em}

\ForAll{$f_k \in F$ (starting from higher to lower priority)} 
	\State \Comment{Relax bandwidth constraint and solve}
	\State Solve $\mathtt{MCP\_HEURISTIC}$($N, s_k, t_k, \mathfrak{D}_k, \widetilde{\mathfrak{B}}_k, D_k, X_k$) by using Algorithm \ref{alg:mcp_org}
    \If {$\mathsf{SolutionFound}$}~~
    \Comment{Path found for $f_k$}
      \State \Comment{Add path to the path vector $\boldsymbol{\mathcal{P}}$}
    \State $\mathcal{P}_k := \mathcal{P}^*$ where $\mathcal{P}^*$ is the solution obtained by Algorithm \ref{alg:mcp_org}
    \Else
    	\State \Comment{Relax delay constraint and try to obtain the path}
    \State Solve $\mathtt{MCP\_HEURISTIC}$($N, s_k, t_k, \widetilde{\mathfrak{D}}_k, \mathfrak{B}_k, X_k, \widehat{B}_k$) by using Algorithm \ref{alg:mcp_org}
        \If {$\mathsf{SolutionFound}$}
        \State \Comment{Path found by relaxing delay constraint}
    \State $\mathcal{P}_k := \mathcal{P}^*$ 
    \Comment{Add path to the path vector}
    \State \Comment{Update remaining available bandwidth} 
    \State $B_e(u,v) := B_e(u,v) - B_k, ~\forall (u,v) \in \mathcal{P}_k$ 
    	\Else
        	\State $\mathcal{P}_k := \mathsf{False}$ ~~\Comment{Unable to find any path for $f_k$}
		\EndIf
    \EndIf
\EndFor

\end{footnotesize}

\end{algorithmic}
\caption{Layout Path Considering Delay and Bandwidth Constraints}
\label{alg:mcp_path_delay_bw}
\end{algorithm}



For each flow $f_k \in F$, starting with highest (\eg the flow with tighter delay requirement) to lowest priority, we first keep the delay constraint unmodified and relax the bandwidth constraint by using Eq. (\ref{eq:bw_relax})  and solve $\mathtt{MCP\_HEURISTIC}$($N, s_k, t_k, \mathfrak{D}_k, \widetilde{\mathfrak{B}}_k, D_k, X_k$) (Line 3) using Algorithm \ref{alg:mcp_org}. If there exists a solution, the corresponding path $\mathcal{P}_k$ is assigned for $f_k$ (Line 6). However, if relaxing bandwidth constraint is unable to return a path, we further relax delay constraint by using Eq. (\ref{eq:delay_relax}), keeping bandwidth constraint unmodified and solve $\mathtt{MCP\_HEURISTIC}$($N, s_k, t_k, \widetilde{\mathfrak{D}}_k, \mathfrak{B}_k, X_k, \widehat{B}_k$) (Line 9). If the path is not found after \textit{both} relaxation steps, the algorithm returns False (Line 14) since it is not possible to assign a path for $f_k$ such that both delay and bandwidth constraints are satisfied. Note that the heuristic solution of the MCP depends of the parameter $X_k$. From our experiments we find that if there exists a solution, the algorithm is able to find a path as long as $X_k \geq 10$.

\subsection{Complexity Analysis}

Note that Line 8 in Algorithm \ref{alg:mcp_org} is executed at most $(C_2 + 1)(V-1)E$ times. Besides, if there exists a path, the worst-case complexity to extract the path is $|\mathcal{P}| C_2$. Therefore, time complexity of Algorithm \ref{alg:mcp_org} is $O(C_2 (V E + |\mathcal{P}|)) = O(C_2 V E)$. Hence the worst-case complexity (\eg when both of the constraints need to be relaxed) to execute Algorithm \ref{alg:mcp_path_delay_bw} for each flow $f_k \in F$ is $O((X_k + \widehat{B}_k )V E)$.

\section{Implementation}
\label{sec:implementation}

We implement our prototype as an {\em application that uses the northbound API}
for the Ryu controller \cite{ryu}. The prototype application accepts the
specification of flows in the SDN. The flow specification contains the
classification, bandwidth requirement and delay budget of each individual flow. In order for a given flow $f_k$ to be realized in the network, the control-plane state of the SDN needs to be modified. The control-plane needs to route traffic along the path calculated for each $f_k$ as described in Section \ref{sec:algo}. In this section, we describe how this is accomplished by decomposing the network-wide state modifications into a set of smaller control actions (called Intents) that occur at each switch. 

%

\begin{figure}
\centering
\vspace*{-1em}
\includegraphics[width=2.75in]{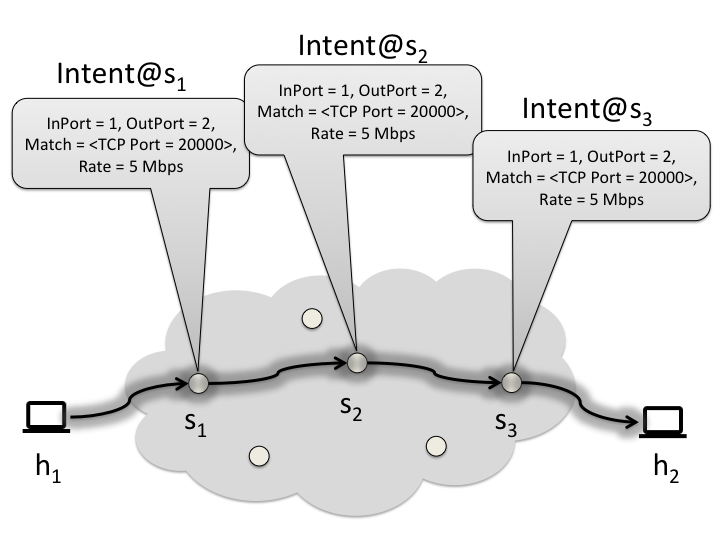}
\caption{Illustration of decomposition of a flow $f_k$ into a set of intents: $f_k$ here is a flow from the source host $h_1$ to the host $h_2$ carrying mission-critical DNP3 packets with destination TCP port set to $20,000$. In this example, each switch that $f_k$ traverses has exactly two ports.}
\label{fig:intents_illustration}
\vspace{-0.2in}
\end{figure} 

\subsection{Forwarding Intent Abstraction}

An \textit{intent} represents the {\em actions performed on a given packet at each
individual switch}. Each flow $f_k$ is decomposed into a set of intents as shown
in Figure \ref{fig:intents_illustration}. The number of intents that are
required to express actions that the network needs to perform (for packets in a
flow) is the same as the number of switches on the flow path. Each
intent is a tuple given by $(\mathsf{Match}, \mathsf{Input Port},
\mathsf{Output Port}, \mathsf{Rate})$. Here, $\mathsf{Match}$ defines the set
of packets that the intent applies to, $\mathsf{Input Port}$ and
$\mathsf{Output Port}$ are where the packet arrives and leaves the switch and
finally, the $\mathsf{Rate}$ is intended data rate for the packets matching the
intent. In our implemented mechanism for laying down flow paths, each intent
translates into a single OpenFlow \cite{specification20131} flow rule that is
installed on the corresponding switch in the flow path. 

\subsection{Bandwidth Allocation for Intents}
In order to guarantee bandwidth allocation for a given flow $f_k$, each one of
its intents (at each switch) in the path need to allocate the same amount of
bandwidth. As described above, each intent maps to a flow rule and the flow
rule can refer to a meter, queue or both. However, meters and queues are
precious resources and not all switch implementations provide both of them.
As mentioned earlier (Section \ref{sec:system_model}), 
we use the strategy of one queue per flow that guarantees 
better isolation among flows and results in stable delays.

\subsection{Intent Realization}
Each intent is realized by installing a corresponding flow rule by using the northbound API of the Ryu controller. Besides using the intent's $\mathsf{Match}$ and $\mathsf{Output Port}$, these flow rules refer to corresponding queue and/or meter. If the meters are used, then they are also synthesized by using the controller API. However, OpenFlow does not support installation of queues in its controller-switch communication protocol, hence the queues are installed separately by interfacing directly with the switches by using a switch API or command line interface. 

\section{Evaluation}
\label{sec:evaluation}










In this section, we evaluate our proposed solutions using the following
methods:
\ca an exploration of the design space/performance of the path layout 
algorithm in Section \ref{subsec:exp:synthetic}, and
\cb an empirical evaluation, using Mininet, that demonstrates the effectiveness
of our end-to-end delay guaranteeing mechanisms even in the presence of other
traffic in the network (Section \ref{subsec:exp:mininet}).
The parameters used in the experiments are summarized in Table \ref{tab:ex_param}.

\begin{table}[!htb]
\vspace{-0.1in}
\renewcommand{\arraystretch}{1.2}
\caption{Experimental Platform and Parameters}
\label{tab:ex_param}
\centering
\begin{small}
\begin{tabular}{p{4.20cm}||p{3.30cm}}
\hline 
\bfseries Artifact/Parameter & \bfseries Values\\
\hline\hline
Number of switches & 5 \\
Bandwidth of links & 10 Mbps \\
Link delay & [25, 125] $\mu$s \\
Bandwidth requirement of a flow & [1, 5] Mbps \\
SDN controller & Ryu 4.7 \\
Switch configuration & Open vSwitch 2.3.0 \\
Network topology & Synthetic/Mininet 2.2.1 \\
OS & Debian, kernel 3.13.0-100\\
\hline
\end{tabular}
\end{small}
\vspace{-0.1in}
\end{table}

\subsection{Performance of the Path Layout Algorithms}
\label{subsec:exp:synthetic}

\subsubsection*{Topology Setup and Parameters} 
\label{sec:synthetic_setup}

In the first set of experiments we explore the design space (\eg feasible delay
requirements) with randomly generated network topologies and synthetic flows. For
each of the experiments we randomly generate a graph with 5 switches and
create $f_k \in [2, 20]$ flows. 
Each switch has 2 hosts connected to it. We
assume that the bandwidth of each of the links $(u,v) \in E$ is 10 Mbps 
(\eg IEEE 802.3t standard \cite{ethernet_standard}). 
The link delays are 
randomly generated within [25, 125] $\mu$s (refer to Appendix \ref{subsec:delay_cal} for the calculation of link delay parameters). For each randomly-generated topology, 
we consider the bandwidth requirement as $B_k \in [1, 5]$ Mbps, $\forall f_k$.

\subsubsection*{Results}

\begin{figure}[!t]
\centering
\vspace*{-2em}
\includegraphics[width=3.3in]{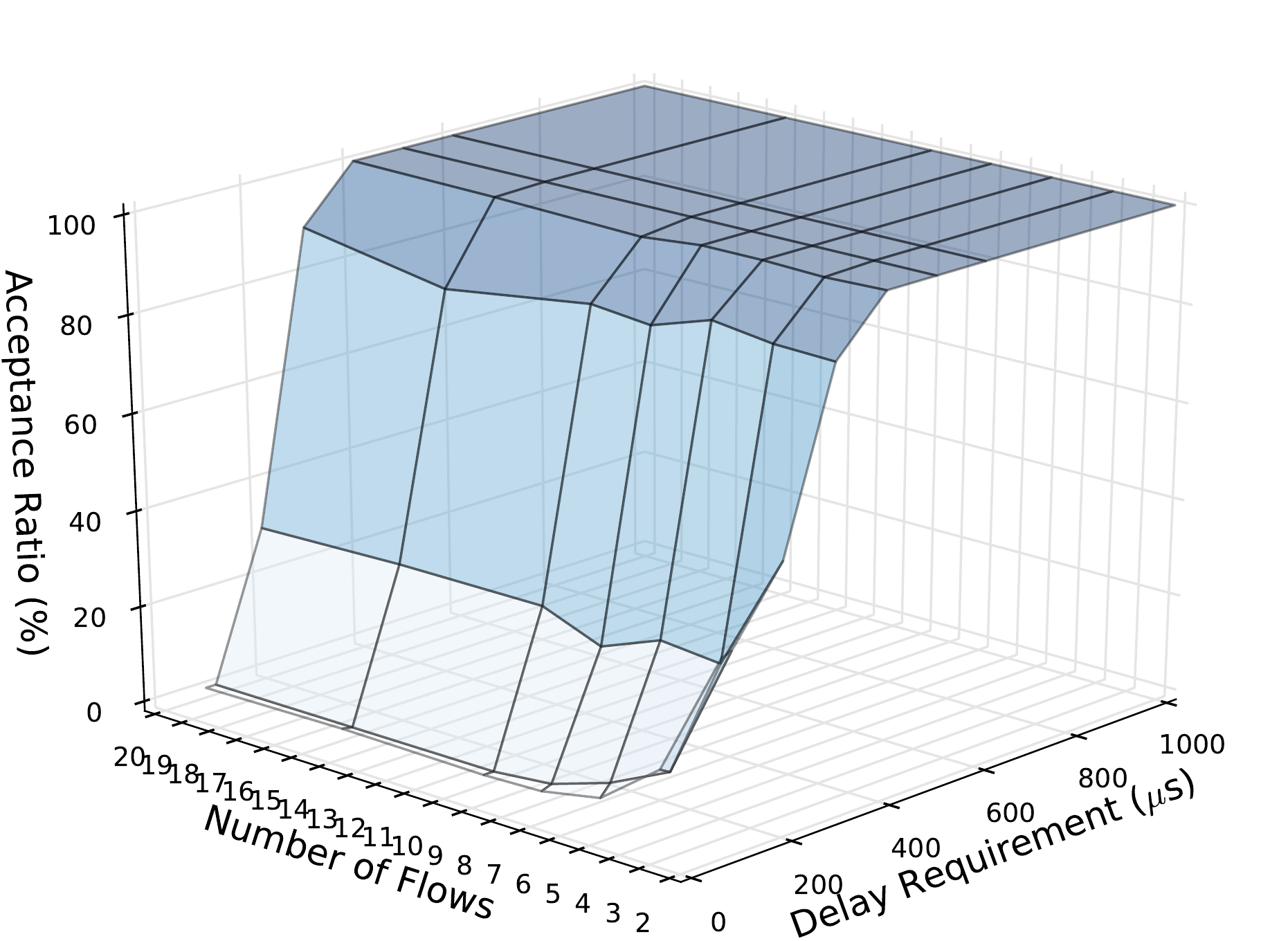}
\caption{Schedulability of the flows in different network topology. For each of the (delay-requirement, number-of-flows) pair (\eg $x$-axis and $y$-axis of the figure), we randomly generate 250 different topology. In other words, total 8 $\times$ 7 $\times$ 250 = 14,000 different topology were tested in the experiments.}
\label{fig:schedulabiliy}
\vspace{-0.2in}
\end{figure}

We say that a given network topology with set of flows is \textit{schedulable}
if all the real-time flows in the network can meet the delay and bandwidth
requirements. We use the \textit{acceptance ratio} metric ($z$-axis in
Fig.~\ref{fig:schedulabiliy}) to evaluate the schedulability of the flows. 
{\em The acceptance ratio is defined as the number of accepted topologies
(\eg the flows that satisfied bandwidth and delay constraints) over the total
number of generated ones}. To observe the impact of delay budgets in different
network topologies, we consider the end-to-end delay requirement $D_k,~ \forall
f_k \in F$ as a function of the topology. In particular, for each randomly
generated network topology $G_i$ we set the minimum delay requirement for the
highest priority flow as $D_{min} = \beta \delta_i$ $\mu$s, and increment it by
$\frac{D_{min}}{10}$ for each of the remaining flows. Here $\delta_i$ is the
diameter (\eg maximum eccentricity of any vertex) of the graph $G_i$ in the
$i$-th spatial realization of the network topology, $\beta =
\frac{D_{min}}{\delta_i}$ and $D_{min}$ represents $x$-axis values of
Fig.~\ref{fig:schedulabiliy}\footnote{Remember our ``delay-monotonic'' priority
assignment where flows with lower end-to-end delays have higher priority.}. 
For each (delay-requirement, number-of-flows) pair, we randomly generate 250 
different topologies and measure the acceptance ratios. As Fig.~\ref{fig:schedulabiliy} 
shows, stricter delay requirements (\eg less than 300 $\mu$s for a set of 20 flows) limit the 
schedulability (\eg only 60\% of the topology is schedulable). Increasing the
number of flows limits the available resources (\eg bandwidth) and thus the
algorithm is unable to find a path that satisfies the delay requirements of \textit{all}
the flows. 

\subsection{Experiment with Mininet Topology: Demonstrating that the End-to-End Delay Mechanisms Work} 
\label{subsec:exp:mininet}

\subsubsection*{Experimental Setup}

The purpose of the experiment is to evaluate whether our controller rules and
queue configurations can provide isolation guarantees so that the real-time flows can meet
their delay requirement in a practical setup. We evaluate the performance of
the proposed scheme using Mininet \cite{lantz2010network} (version 2.2.1) where switches are
configured using Open vSwitch \cite{ovs} (version 2.3.0). We use Ryu \cite{ryu}
(version 4.7) as our SDN controller. For each of the experiments we randomly
generate a Mininet topology using the parameters described in
Table \ref{tab:ex_param}.
We develop flow rules in the queues to enable connectivity among hosts in
different switches. The packets belonging to the real-time flows are queued
separately in individual queues and each of the queues are configured at a
maximum rate of $B_k \in [1, 5]$ Mbps. If the host exceeds the configured
maximum rate of $B_k$, our ingress policing throttles the traffic before it
enters the switch\footnote{In real
systems, the bandwidths allocation would be overprovisioned (as mentioned earlier), our evaluation takes a conservative approach.}. 

To measure the effectiveness of our prototype with mixed
(\eg real-time and non-critical) flows, we enable [1,3]
non-critical flows in the network.  All of the low-criticality
flows use a {\em separate, single queue} and are served in a FIFO manner -- it is the
``default'' queue in OVS. Since many commercial switches (\eg Pica8 P-3297, HPE FlexFabric 12900E, \etc) supports up to 8 queues, in our Mininet experiments we limit the maximum number of real-time flows to 7 (each uses a separate queue) and use the remaining 8th queue for non-critical flows. 
Our flow rules isolate the
non-critical flows from real-time flows. All the experiments
are performed in an Intel Xeon 2.40 GHz CPU and Linux kernel version 3.13.0-100.

We use \texttt{netperf} (version 2.7.0)\cite{netperf} to generate the UDP traffic\footnote{Remember
that most hard real-time systems use UDP traffic \cite{rt_sdn_kaist, afdx_evolution}.} between the
source and destination for any flow $f_k$. Once the flow rules and queues are
configured, we triggered packets starting at the source $s_k$ destined to
host $t_k$ for each of the flows $f_k$. The packets are sent at a burst of 5
with 1 ms inter burst time. All packet flows are triggered simultaneously and
last for 10 seconds.

  
We assume flows are indexed based on priority, \ie $D_1 < D_2 < \cdots <
D_{|F|}$ and 
randomly generate 25
different network topologies. We set $D_1 = 100 \delta_i$ $\mu$s and increment with 10 for each of
the flow $f_k \in F, k>1$ where $\delta_i$ is the diameter of the graph $G_i$ in the $i$-th spatial
realization of the network topology. For each topology, we
randomly generate the traffic with required bandwidth $B_k \in [1, 5]$ Mbps and
send packets between source ($s_k$) and destination ($t_k$) hosts for 5 times
(each transmission lasts for 10 seconds) and log the worst-case round-trip
delay experienced by any flow.



\subsubsection*{Experience and Evaluation}

\begin{figure}[t!]
\addtolength{\subfigcapskip}{-0.1in}
\centering
\subfigure[]{\includegraphics[width=\linewidth]{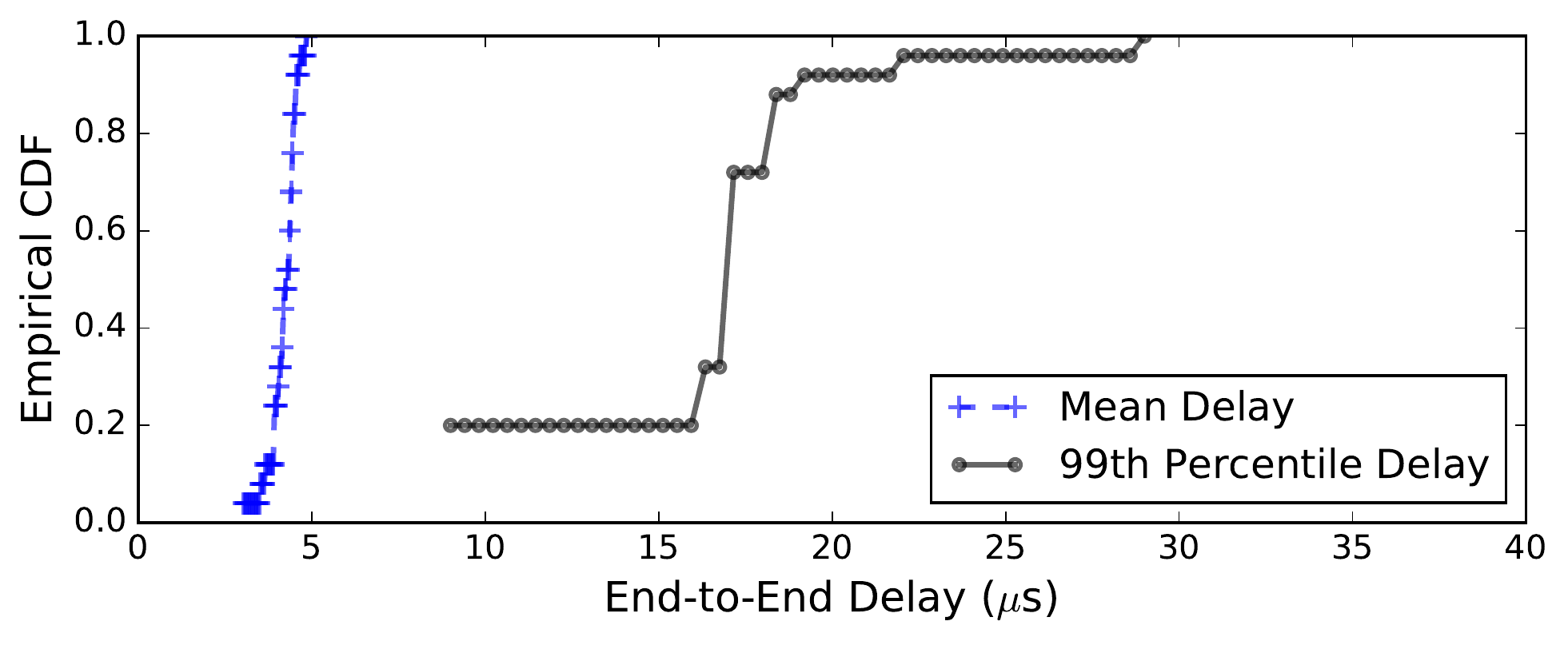}
\label{fig:mhasan_mean_99p_delay}}

\vspace*{-1.0em}
\subfigure[]{\includegraphics[width=\linewidth]{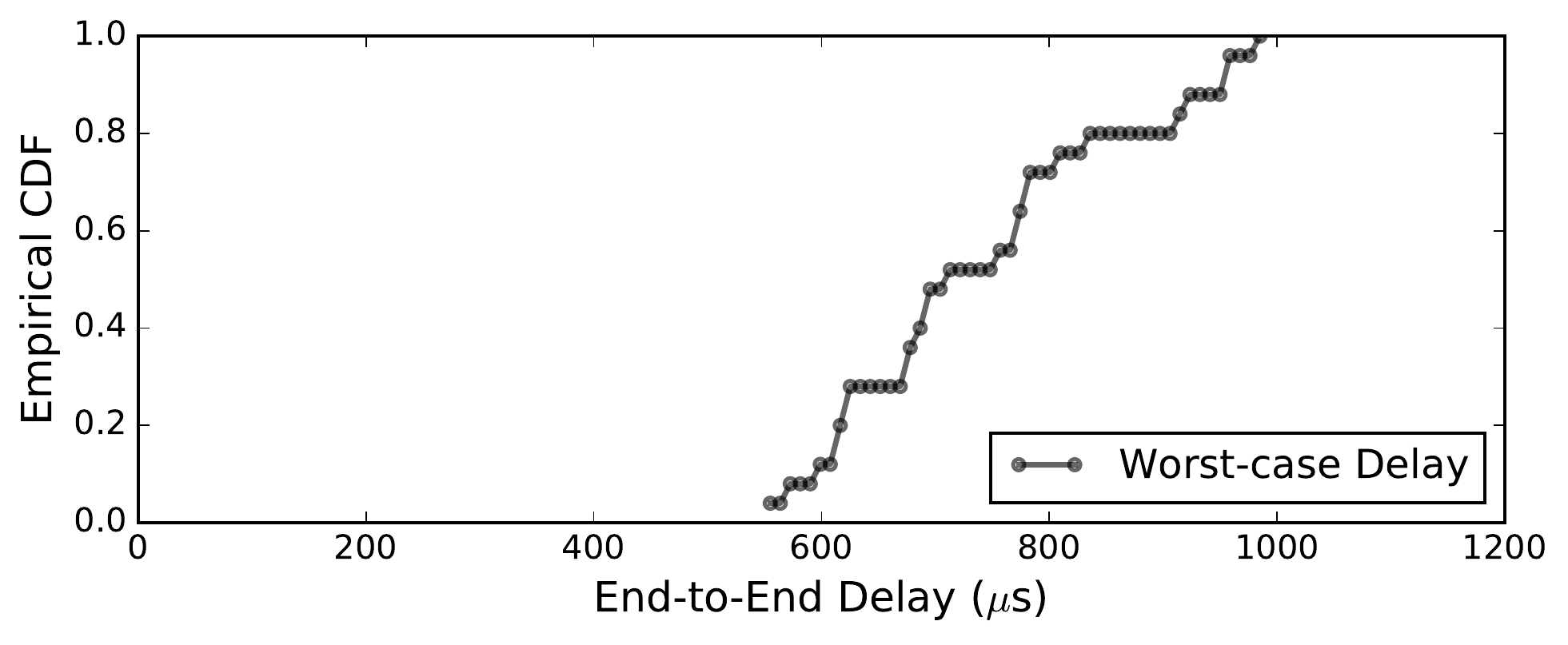}\label{fig:mhasan_max_delay}}
\caption{(a) The empirical CDF of: (a) average and $99^{th}$ percentile, (b)  worst-case round-trip delay. 
We set the number of flows $f_k = 7$ and
examine 7 $\times$ 25 $\times$ 5 packet flows (each for 10 seconds) to obtain the experimental traces.}
\label{fig:mhasan_delay}
\vspace{-0.2in}
\end{figure}

In Fig. \ref{fig:mhasan_delay} we observe the impact of number of flows on the
delay. Experimental results are illustrated for the schedulable flows (\viz the
set of flows for which \textit{both} delay and bandwidth constraints are
satisfied). 

The y-axis of Fig. \ref{fig:mhasan_delay} represents the empirical CDF of
average/$99^{th}$ percentile (Fig.
\ref{fig:mhasan_mean_99p_delay}) and worst-case (Fig. \ref{fig:mhasan_max_delay})
round-trip delay experienced by any flow. The empirical CDF is defined as
$G_\alpha(\jmath) = \frac{1}{\alpha} \sum_{i=1}^\alpha \mathbb{I}_{[\zeta_i
\leq \jmath]}$, where $\alpha$ is the total number of experimental
observations, ${\zeta}_i$ round-trip delay the $i$-th experimental observation,
and $\jmath$  represents the $x$-axis values (\viz round-trip delay) in Fig.
\ref{fig:mhasan_delay}. The indicator function $\mathbb{I}_{[\cdot]}$ outputs
1 if the condition $[\cdot]$ is satisfied and 0 otherwise. 

From our experiments we find that, the non-critical flows \textit{do not}
affect the delay experienced by the real-time flows and the average as well as the $99^{th}$
percentile delay experienced by the real-time flows \textit{always} meet their delay
requirements. This is because our flow rules and queue configurations isolate
the real-time flows from the non-critical traffic to ensure that the end-to-end
delay requirements are satisfied. We define the \textit{expected delay bound}
as the expected delay if the packets are routed through the diameter (\ie the
greatest distance between any pair of hosts) of the topology and given by
$\mathfrak{D}_i(u, v) \times  \delta_i$ and bounded by $[25 \delta_i, 125
\delta_i]$ where $\mathfrak{D}_i(u, v) \in [25, 125]$ is the delay between the
link $(u,v)$ in $i$-th network realization. As seen in Fig.
\ref{fig:mhasan_mean_99p_delay}, the average and
$99^{th}$ percentile round-trip delay are significantly less than the minimum
expected round-trip delay bound (\eg 2 $\times$ 25 $\times$ 4 = 200 $\mu$s).
This also validates the effectiveness of Algorithm
\ref{alg:mcp_path_delay_bw}. Besides, as seen in Fig.
\ref{fig:mhasan_delay}, the worst-case delay is also less than the maximum
expected delay bound (\eg 1000 $\mu$s) with probability 1.

\begin{figure}[!t]
\centering
\includegraphics[width=\linewidth]{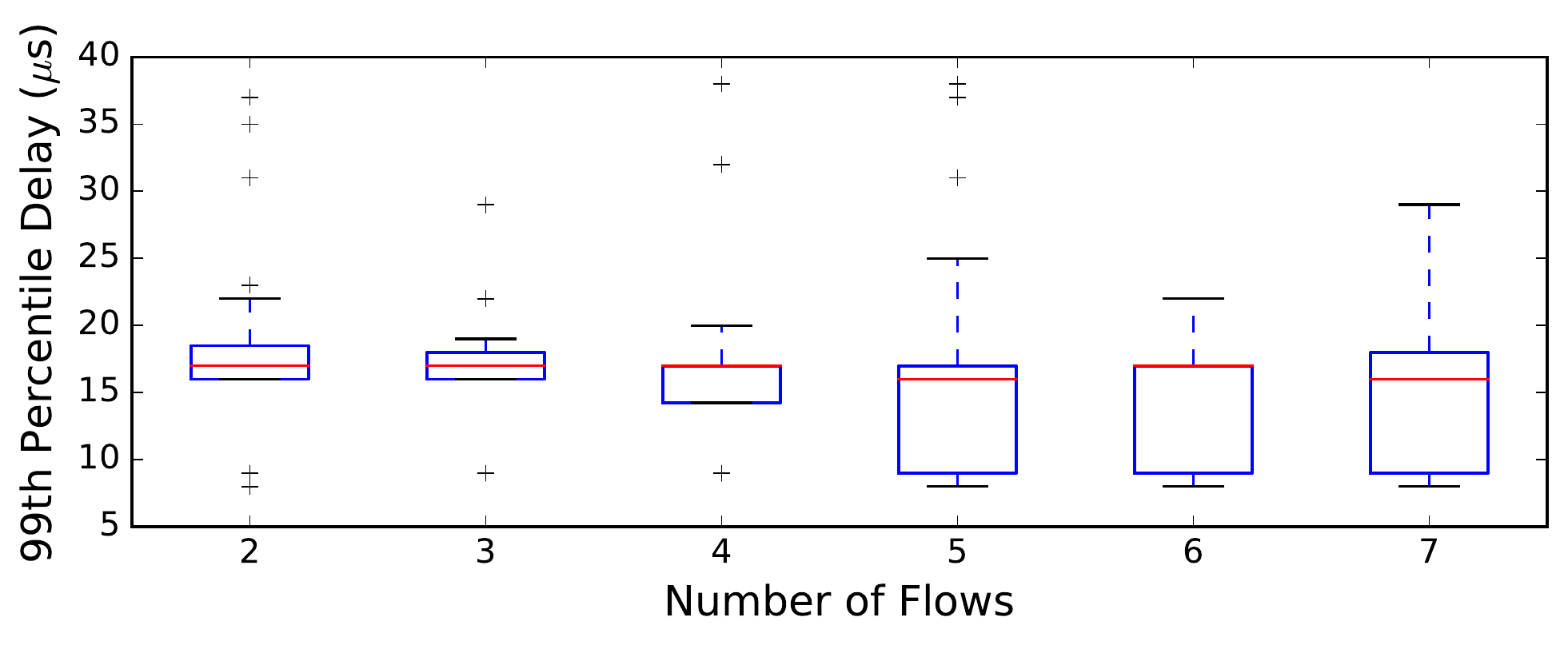}\caption{End-to-end round-trip $99^{th}$ percentile delay with varying number of flows. For each set of flow $f_k \in [2, 7]$, we examine $f_k \times 25 \times 5$ packet flows (each for 10 seconds).
}
\label{fig:mhasan_delay_box}
\vspace{-0.1in}
\end{figure}

In Fig. \ref{fig:mhasan_delay_box} we illustrate the $99^{th}$ percentile round-trip
delay (represents the y-axis in the figure) with different number of flows
(x-axis). Recall that in our experimental setup we assume at most 8 queues are available in the switches where 7 real-time flows are assigned to each of 7 queues and the other queue is used for [1, 3] non-critical flows. As shown in Fig. \ref{fig:mhasan_delay_box}, increasing the number of flows slightly decreases quality of
experience (in terms of end-to-end delays). With increasing
number of packet flows the switches are simultaneously processing forwarding
rules received from the controller -- hence, it increases the round-trip delay. Recall
that the packets of a flow are sent in a bursty manner using \texttt{netperf}.
Increasing number of flows in the Mininet topology increases the packet loss and
thus causes higher delay. 

For our experiments with Mininet and \texttt{netperf} generated traffic, we do
\textit{not} observe any instance for which a set of schedulable flow misses its
deadline (\ie packets arriving {\em after} the passing of their end-to-end delay
requirements). Thus, based on our empirical results and the constraints provided to the path layout algorithm, we can assert that the schedulable real-time flows will meet their corresponding end-to-end delay requirements.


\section{Discussion}
\label{subsec:limitations}

Despite the fact that we provide an initial approach to leverage the benefits of the SDN architecture to guarantee guarantee end-to-end delay in safety-critical hard RTS, our proposed scheme has some limitations and can be extended in several directions. To start with, 
most hardware switches limit the maximum number of individual queues\footnote{\eg Pica8 P-3297 and HPE FlexFabric 12900E switches support at most 8 queues.} that can be allocated to flows. Our current intent realization mechanism reserves one queue per port for each Class I flow. This leads to depletion of available queues. Hence, we need smarter methods to \textit{multiplex} Class I flows through limited resources and yet meet their timing requirements. Our future work will focus on developing sophisticated schemes for ingress/egress filtering at each RT-SDN-enabled switch. This will also help us better identify the properties of each flow (priority, class, delay, \etc) and then develop scheduling algorithms to meet their requirements. 

In this work we allocate separate queues for each flow and layout paths based on the ``delay-monotonic'' policy. However establishing and maintaining the flow priority is \textit{not} straightforward if the ingress policing requires to share queues and ports in the switches. Many existing mechanisms to enforce priority are available in software switches (\eg the hierarchical token buckets (HTB) in Linux networking stack). In our experience, enabling priority on hardware switches has proven difficult due to firmware bugs.


Finally, we do not impose any admission control policy for the unschedulable (\ie the flows for which the delay and bandwidth constraints are not satisfied) flows. One approach to enable admission control is to allow $m$ out of $k$ ($m < k$) packets of a low-priority flow to meet the delay budget by leveraging the concept of $(m,k)$ scheduling \cite{m_k_overload} in traditional RTS.

\section{Related Work}
\label{sec:related_work}




There have been several efforts to study the provisioning a network such that
it meets bandwidth and/or delay constraints for the traffic flows. Results from
the network calculus (NC) \cite{le2001network} framework offer a concrete way
to model the various abstract entities and their properties in a computer
network. NC-based models, on the other hand, do not prescribe any formulation
of flows that meet given delay and bandwidth guarantees. For synthesis, the
NP-complete MCP comes close and Shingang \etal formulated a heuristic algorithm
\cite{mcp_klara} for solving MCP. We model our delay and bandwidth constraints
based on their approach. 

There are recent standardization efforts
such as IEEE 802.11Qbv \cite{ieee80211qbv} which aim to codify best practices for provisioning QoS using Ethernet. These approaches focus entirely on meeting guarantees and do not attempt to optimize link bandwidth.  However, the global view of the network provided by the SDN architecture allows us to optimize path layouts by formulating it as an MCP problem.

There have been some prior attempts at provisioning SDN with worst-case delay
and bandwidth guarantees.  Azodolmolky \etal proposed a NC-based model
\cite{azodolmolky2013analytical} for a single SDN switch that provides an upper
bound on delays experienced by packets as they cross through the switch. Guck
\etal used mixed integer program (MIP) based formulation
\cite{guck2014achieving} for provisioning end-to-end flows with delay
guarantees -- they do not provide a solution of what traffic arrival rate to
allocate for queues on individual switches for a given end-to-end flow. 

A QoS-enabled management framework to allow end-to-end communication over SDN is proposed in literature \cite{Xu2015}. 
It uses flow priority and queue mechanism to obtain QoS control to satisfy the requirement  but did not demonstrate schedulability under different constraints.  A scalable routing scheme was developed in literature \cite{rt_sdn_kaist} that re-configures existing paths and calculates new paths based on the global view and bandwidth guarantees. The authors also present a priority ordering scheme to avoid contention among flows sharing the same switch. However, the basic requirement of the model used in that work (\ie end-to-end delay being less than or equal to minimum separation times between two consecutive messages) limits applicability of their scheme for a wide
range of applications. 

Avionics full-duplex switched Ethernet
(AFDX)~\cite{ARINC2009,land2009,Charara2006} is a deterministic data network
developed by Airbus for safety critical applications. 
The switches in AFDX architecture are interconnected using full duplex links, and static paths with
predefined flows that pass through network are set up. 
Though such solutions aim to
provide deterministic QoS guarantees through static routing, reservation and isolation, they
impose several limitations on optimizing the path layouts and on different
traffic flows. 
There have been studies towards evaluating the upper bound on the end-to-end
delays in AFDX networks\cite{Charara2006}. The evaluation seems to depend on the AFDX
parameters though.


There are several protocols proposed in automotive communication networks
such as controller area network (CAN) \cite{can2014} and FlexRay \cite{flex2016}. 
These protocols are designed to provide strong real-time guarantees but have limitations in how to extend it to varied network lengths, different traffic flows and complex network topologies. With SDN architectures and a flexible QoS framework 
proposed in this paper, one could easily configure COTS components and
meet QoS guarantees with optimized path layouts. 

Heine \etal proposed a design and built a real-time middleware system, CONES
(COnverged NEtworks for SCADA) \cite{Heine2011} 
that enables the communication of data/information in SCADA
applications over single physical integrated networks. 
However, the authors did not explore the synthesis of
rules or path optimizations based on bandwidth-delay requirements -- all of which are carried out by our system.
Qian \etal implemented a hybrid EDF packet scheduler \cite{Qian2015} for
real-time distributed systems. The authors proposed a proportional bandwidth sharing strategy based on number
of tasks on a node and duration of these task, due to partial information of the network.
In contrast, the SDN
controller has a global view of the network, thus allowing for more flexibility to
synthesize and layouts the paths and more control on the traffic.  

The problem of end-to-end delay bounding in RTS is addressed in literature~\cite{Jin2013}. The authors choose avionics systems composed of end devices,
and perform timing analysis of the delays introduced by end points and the
switches. However, the proposed approach requires modification to the
switches. Besides the authors do not consider the bandwidth limitations,
variable number of flows and flow classifications. 

There is a lot of work in the field of traditional real-time networking (too many
to enumerate here) but the focus on SDN is what differentiates our work.

\section{Conclusion}
\label{sec:concl}

With the proliferation of commercial-off-the-shelf (COTS) components, 
designers are exploring new ways of using them, even in critical systems
(such as RTS).
Hence, there is a need to understand the inherent trade-offs (less customization)
and advantages (lower cost, scalability, better support and more choices)
of using COTS components in the design of such systems. In this paper, we
presented mechanisms that provide end-to-end delays for critical traffic
in real-time systems using COTS SDN switches. Hence, future RTS can be
better managed, less complex (fewer network components to deal with) and
more cost effective.

\bibliographystyle{abbrv}
\bibliography{securecore,sibin.security,sibin.sdn,other,monowar,konstantin}  

\appendices

\section{Delay Calculations}
\label{subsec:delay_cal}


Remember that some of the critical pieces of information that is required for
any such scheme (for ensuring end-to-end delays) is a measure of the delays
imposed by the various components in the system. Hence, we need to obtain
network delays at each link. We use these estimated delays as the weights of
edges of the network graph in the MCP algorithm within the experimental setup to
obtain solutions. As discussed earlier, we assume zero queuing delay. The
transmission and propagation delays are a function of the physical properties
of the network topology. However, the processing delay of an individual switch
for a single packet can be empirically obtained. Here we describe our method to
obtain upper-bounds on each of these delay components.

\subsection*{Estimation of Propagation Delay}
The transmission delay is calculated as $\frac{\text{packet
length}}{\text{bandwidth allocated}}$. In our experiments we assume the packet
length is [25, 125] bytes and the maximum bandwidth can be allocated allocated
in a specific link is 10 Mbps. Then transmission delay on that link will be
upper bounded by $\frac{125 \times 8~\text{bits}}{10~ \text{Mbps}}$ = 100
$\mu$s. Therefore delay of the edge, \ie $\mathfrak{D}_k(u, v)$, $\forall (u,v)
\in E$ is upper bounded by 3.6+0.505+100 $\approx$ 105 $\mu$s.

\subsection*{Estimation of Transmission Delay}
The propagation delay depends on the physical link length and propagation speed
in the medium. In the physical media, the speed varies .59$c$ to .77$c$ \cite{prop_delay} where
$c$ is speed of light in vacuum. We assume that the length of any link in the
network to be no more that 100 $m$. Therefore the propagation delay is upper bounded
by $\frac{100 \rm m}{0.66 \times 3 \times 10^6} = 505$ ns in fiber-link media. 

\subsection*{Estimation of Processing Delays}

We experimented with a software switch, Open vSwitch (OVS) \cite{ovs} version
2.5.90 to compute the time it takes to process a packet within its data path.
Since this timing information is platform/architecture dependent, we summarized
the hardware information of our experimental platform in Table
\ref{tab:delay_param}. 

\begin{table}[!htb]
\renewcommand{\arraystretch}{1.2}
\caption{Hardware used in Timing Experiments}
\label{tab:delay_param}
\centering
\begin{small}
\begin{tabular}{p{3.20cm}||p{4.30cm}}
\hline 
\bfseries Artifact & \bfseries Info\\
\hline\hline
Architecture &              i686 \\
CPU op-modes &       32-bit, 64-bit \\
Number of CPUs &                     4 \\
Threads per core &     2 \\
Cores per socket &     2 \\
CPU family &               6 \\
L1d and L1i cache &              32K \\
L2 and L3 cache &               256K and 3072K, respectively \\
\hline
\end{tabular}
\end{small}
\end{table}

\begin{figure}[H]
\centering
\includegraphics[width=2.7in]{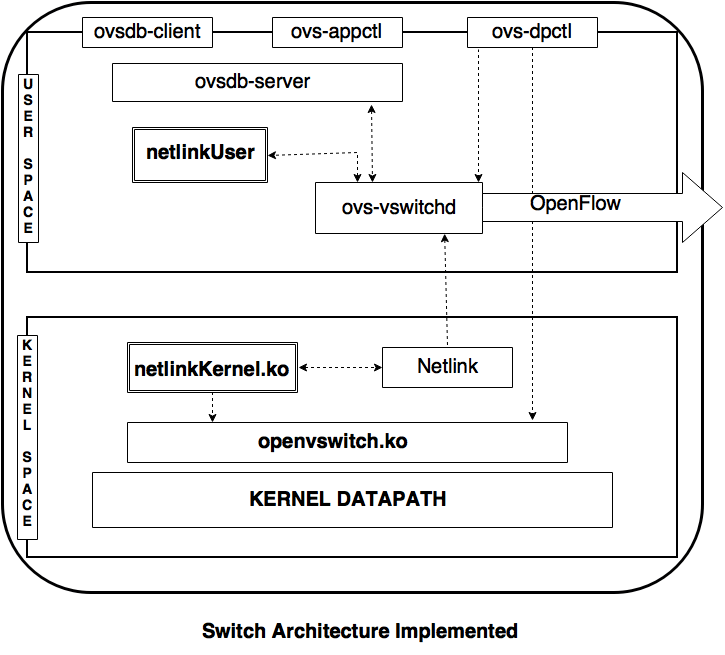}
\caption{Interaction of kernel timing module with the existing OVS architecture.}
\label{fig:ovs-arch-mod}
\end{figure}

We modified the kernel-based OVS data path module called
\texttt{openvswitch.ko} to measure the time it takes for a packet to move from
an ingress port to an egress port. We used \texttt{getnstimeofday()} for
high-precision measurements. We also developed a kernel module called
\texttt{netlinkKernel.ko} that copies the shared timing measurement data
structure between the two kernel modules and communicates it with a user space
program called \texttt{netlinkUser}. We disabled scheduler preemptions in the
\texttt{openvswitch.ko} by using the system calls \texttt{get\_cpu()} and
\texttt{put\_cpu()}, hence the actual switching of the packets in the data path
is not interfered by the asynchronous communication of these measurements by
\texttt{netlinkKernel.ko}. We also used compilation flags to ensure that
\texttt{openvswitch.ko} always executes on a specified, separate, processor
core of its own (with no interference from any other processes, both from the
user space or the operating system). For fairness in the timing measurements
and stabilized output, we disabled some of the Linux background processes (\eg
SSH server, X server) and built-in features (\eg CPU frequency scaling). Figure
\ref{fig:ovs-arch-mod} illustrates the interaction between the modified kernel
data path and our user space program.

We used the setup described above with Mininet and Ryu Controller. We evaluated
the performance and behavior of OVS data path under different flows, network
typologies and packet sizes. We executed several runs of the experiment with UDP
traffic with different packet sizes (100, 1000, 1600 bytes). We observed that
average processing time for a single packet within the software switch lies
between 3.2 $\mu$s to 4.1 $\mu$s with average being 3.6 $\mu $s and standard
deviation being 329.61 ns. These were the values that were used in the path 
allocation calculations.

\section{Queue Assignment Strategies: Mininet Observations}\label{appsec:mininet_simple_topo}

\begin{figure}[!h]
\centering
\includegraphics[width=3.5in]{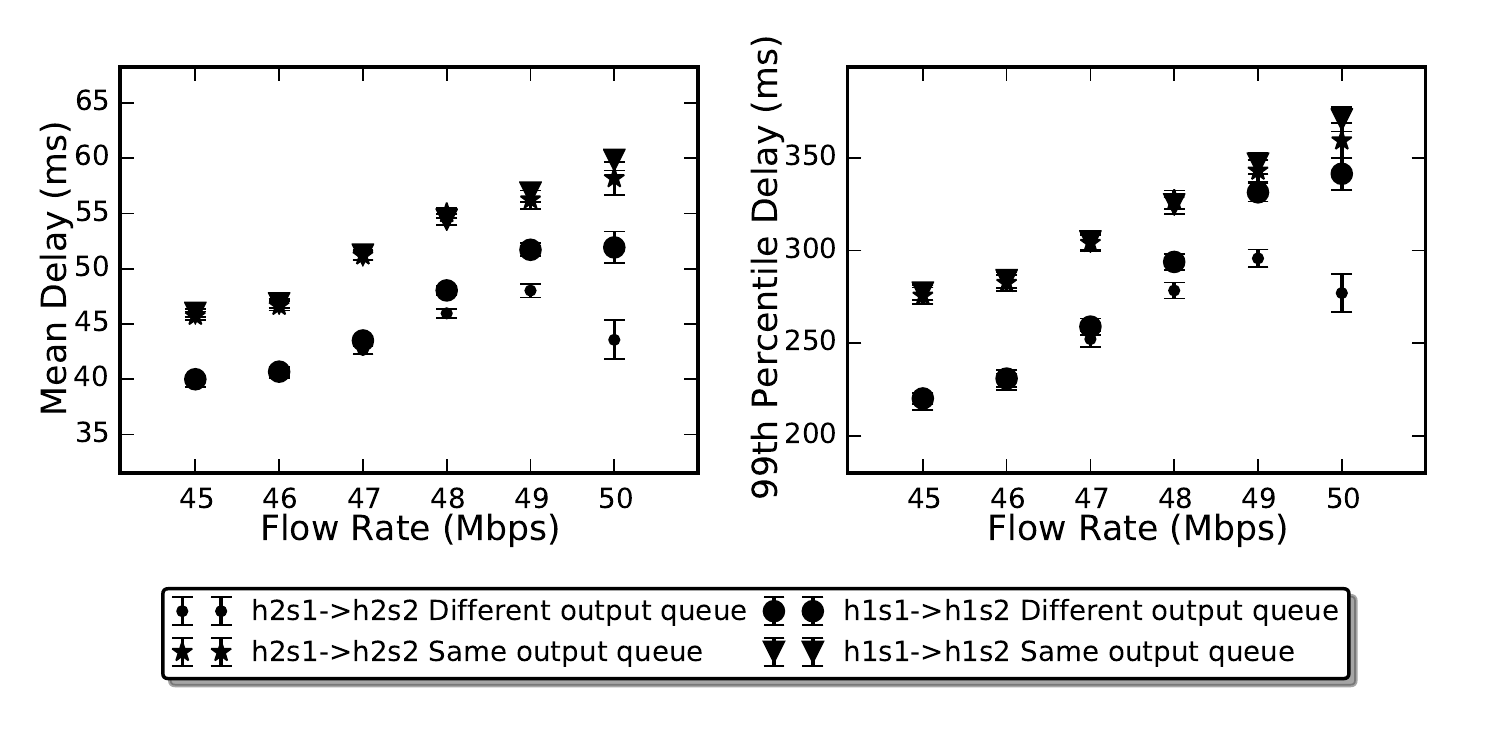}
\caption{The mean and $99^{th}$ percentile per-packet delay for the packets in the active flows in 25 iterations using a two-host four-switch (see Fig. \ref{fig:delay_measurements}) Mininet topology.}
\label{fig:mininet_delay_measurements}
\end{figure}
We also perform experiments with a two switch, four host topology similar that of presented in Section \ref{sec:motivating_example} using Mininet. The purpose of this experiment is to observe the performance impact on software simulations (\eg Mininet topologies) over the actual ones  (hardware switches and ARM hosts).
As we can see in Fig. \ref{fig:mininet_delay_measurements} the trends (\eg isolating flows using separate queues results in lower
delays) are similar in both Mininet and hardware experiments -- albeit the latencies are higher due to it being a software simulation and also affected by other artifacts (\eg the experiments are involved in generating traffic on the same machine).

\end{document}